\documentstyle[twoside,fleqn,espcrc2,epsf]{article}

\newcommand{\nn}{\nonumber}

\newcommand{\lw}[1]{\smash{\lower2.ex\hbox{#1}}}
\def\simlt{\rlap{\lower 3.5 pt\hbox{$\mathchar \sim$}}\raise 1pt \hbox {$<$}}
\def\simgt{\rlap{\lower 3.5 pt\hbox{$\mathchar \sim$}}\raise 1pt \hbox {$>$}}

\newcommand{\VEV}[3]{\left\langle #1\left| #2 \right| #3\right\rangle}

\title{
Domain-wall fermions and chiral symmetries 
}

\author{
Sinya Aoki\address{
Institute of Physics, University of Tsukuba, Tsukuba,
Ibaraki 305-8571, Japan}
}

\begin{document}
\begin{abstract}
We investigate chiral properties of the domain-wall fermion (DWF)
system. After a brief introduction for the DWF,
we summarize the recent numerical results on the chiral properties
of the domain-wall QCD (DWQCD), which seem mutually inconsistent.
We next derive a formula which connects a chiral symmetry
breaking term in the five dimensional DWF Ward-Takahashi identity with
the four-dimensional hermitian Wilson-Dirac operator.
Based on this formula, we propose a solution, which
resolves the inconsistency among recent numerical data,
and give a consistent picture of the chiral properties of the DWQCD.
\end{abstract}

\maketitle

\section{Introduction}

A suitable definition of the chiral symmetry has been a long standing 
problem in lattice field theories. Recently an ultimate solution to this 
problem seems to appear in the form of the Ginsberg-Wilson relation
\cite{GW,Luscher98}.
Two explicit examples of the lattice fermion operators which satisfies the 
Ginsberg-Wilson relation have been found so far:
One is the perfect lattice Dirac operator constructed via the renormalization
group transformation\cite{perfect1} and the other is the
overlap Dirac operator\cite{Neuberger98}
derived from the overlap formalism\cite{NN94}
or from the domain-wall fermion(DWF)\cite{Kaplan92,Shamir93} in
the limit of the infinite length of the 5th dimension.
Since the explicit form is simpler for the latter, a lot of numerical
investigations\cite{Blum-Soni,AIKT,Blum98,cppacs-dwf}
as well as analytic considerations\cite{HJL98,KN99}
have been carried out
for the domain-wall fermion or the overlap Dirac fermion.

In this report we review the recent numerical investigations on 
the chiral properties of the DWF,
which, however, bring puzzling results.
To resolve these puzzles, we derive a formula which connects a chiral symmetry
breaking term with the 4-dimensional hermitian Wilson-Dirac operator.
Using this formula and the eigenvalue distribution of the  
Wilson-Dirac operator, we propose a consistent interpretation
of puzzling results.

\section{Domain-wall fermions} 
Throughout this report we employ a Shamir's variant of
the original domain-wall fermion action\cite{Shamir93,Shamir95},
which is equivalent to the 5-dimensional massive Wilson fermion
with the free boundary condition in the 5-th direction,
interacting with the 4-dimensional gauge field.
Symbolically the action is given by
\begin{eqnarray}
S_F &=& \bar\psi \left[ \gamma^\mu D_\mu (A) +\gamma^5 \partial_5\right]\psi
\nn \\
&+& \bar\psi\left[ D_\mu(A)^2 + \partial_5^2\right]\psi
+ M \bar\psi\psi .
\end{eqnarray}
Note that 
the sign of the mass term $M$ is opposite to 
the usual Wilson fermion. 
Since gauge fields are 4-dimensional, the 5-th coordinate $s$ can
be interpreted as ``flavors'', so that the action can be rewritten
in the 4-dimensional form:
\begin{eqnarray}
S_F &=& \displaystyle\frac{1}{2}\bar\psi_{n, s} \gamma^\mu \left[ 
U_{n,\mu}\psi_{n+\mu, s}- U_{n-\mu,\mu}^\dagger\psi_{n-\mu, s}
\right] \nn \\
&+& \bar\psi_{n, s} \left[ {\cal M} P_R 
+ {\cal M}^\dagger P_L
\right]_{s,t}^{n,m}
\psi_{m, t},
\end{eqnarray}
where the last term is considered as the flavor mixing ``mass'' term.

Now we consider the fermion spectrum derived from this action
for the free theory.
In this case the action in the momentum space becomes
\begin{eqnarray}
S_F &=& \int d^4 p [ \bar\psi_s i \gamma^\mu \sin (p_\mu a) \psi_s 
\nn \\
&+&
\bar\psi_s \left( {\cal M} P_R + {\cal M}^\dagger P_L\right)_{st}\psi_t
]
\end{eqnarray}
where
\[
{\cal M} = \left(
\begin{array}{ccccc}
-W & 1 &       &       &       \\
   &-W & 1     &       &        \\
   &   &\ddots &\ddots &     \\
   &   &       & -W    & 1   \\
   &   &       &       & - W \\
\end{array}
\right) 
\]
\[
{\cal M}^\dagger = \left(
\begin{array}{ccccc}
-W &   &       &       &  \\
 1 &-W &       &       &  \\
   & 1 &\ddots &       &     \\
   &   &\ddots & -W    &     \\
   &   &       &  1    & - W \\
\end{array}
\right) 
\]
with $ W = 1 - M +\sum_\mu [ 1- \cos (p_\mu a)] $.

\begin{table}[bt]
\caption{Relation between $M$ and number of poles}
\label{tab:pole}
\vspace{-0.3cm}
\begin{center}
\begin{tabular}{|l|l|l|}
\hline
condition & $p_\mu$ & \# of poles \\
\hline
$ 0 < M < 2 $ & $(0,0,0,0)$ & 1 \\
$ 2 < M < 4 $ & $(\pi/a,0,0,0)$ & 4 \\
$ 4 < M < 6 $ & $(\pi/a,\pi/a,0,0)$ & 6 \\
$ 6 < M < 8 $ & $(\pi/a,\pi/a,\pi/a,0)$ & 4 \\
$ 8 < M < 10 $ & $(\pi/a,\pi/a,\pi/a,\pi/a)$ & 1 \\
\hline
\end{tabular}
\end{center}
\vspace{-1.0cm}
\end{table}

By setting $ \psi (p)_s = P_R u_R(p)_s+P_L u_L(p)_s$,
one observe that the solution to the equation $ {\cal M}_{st} u_R(p)_t = 0 $ 
($ {\cal M}_{st}^\dagger u_L(p)_t = 0 $)
gives a right-handed (left-handed) zero mode.
The solutions to these equations are given by
\begin{eqnarray}
u_R(p)_s &=& W(p)^{s-1} c_R \\
u_L(p)_s &=& W(p)^{N_s-s} c_L ,
\end{eqnarray}
where $N_s$ is the number of sites in the 5-th dimension.
In order to satisfy the boundary conditions
\begin{eqnarray}
u_R(N_s) &=& W(p)^{N_s-1} c_R = 0 \\
u_L(1) &=& W(p)^{N_s-1} c_L = 0 
\end{eqnarray}
for the above solutions,
the $N_s\rightarrow\infty$ limit is necessary,
together with the condition that $\vert W(p) \vert < 1$, which 
is equivalent to
\begin{eqnarray}
0 < M +\sum_\mu (\cos(p_\mu a)-1) < 2 .
\label{eq:condW}
\end{eqnarray}
Since we have the kinetic term $\gamma^\mu \sin (p_\mu a)$ for the fermion,
the condition that $p_\mu a \simeq 0$ or $\pi$ is required for the existence
of the massless pole in the continuum limit.
Eq.(~\ref{eq:condW}) gives a condition for $M$ and $p_\mu$, which is summarized
in table~\ref{tab:pole}.
It is clear that the value of $M$ (Wilson fermion mass) controls
the number of massless fermion poles.
For $ 0 < M < 2$ we have one physical massless Dirac fermion.

The form of zero-modes obtained above
is too complicated in the coordinate space even for the free theory
to define the quark field as $P_R u_R+P_L u_L$.
Instead we define the quark field as
\begin{eqnarray}
q_n & \equiv & P_R \psi_{n, 1}+ P_L \psi_{n, N_s} \\
\bar q_n & \equiv & \bar\psi_{n, 1} P_L + \bar \psi_{n, N_s}
P_R .
\end{eqnarray}
This definition is reasonable 
since the right-handed zero mode has a peak at $s=1$ and
the left-handed one at $s=N_s$.

According to this definition, the quark mass term is given by
$ - m_f \bar q_n q_n $,
which leads to the non-zero pole mass of the fermion
for $0 < M < 2$:
\begin{eqnarray}
m_{\rm pole} &=& M (2-M) [ m_f + (1-M)^{N_s}].
\end{eqnarray}
The formula tells us that
$ m_{\rm pole} \sim (1-M)^{N_s} \not=0$ even at $m_f =0$ for
$N_s\not=\infty$, corresponding to the fact that no solution
to zero modes exist for finite $N_s$.
On the other hand the fact that $m_{\rm pole} \propto m_f$ in 
the $ N_s\rightarrow\infty$ limit strongly suggests that
the chiral symmetry is realized for the DWF.
The lattice QCD with the DWF is called the domain-wall QCD (DWQCD).
\begin{figure}[tb]
\vspace{-0.5cm}
\centerline{\epsfxsize=6.0cm \epsfbox{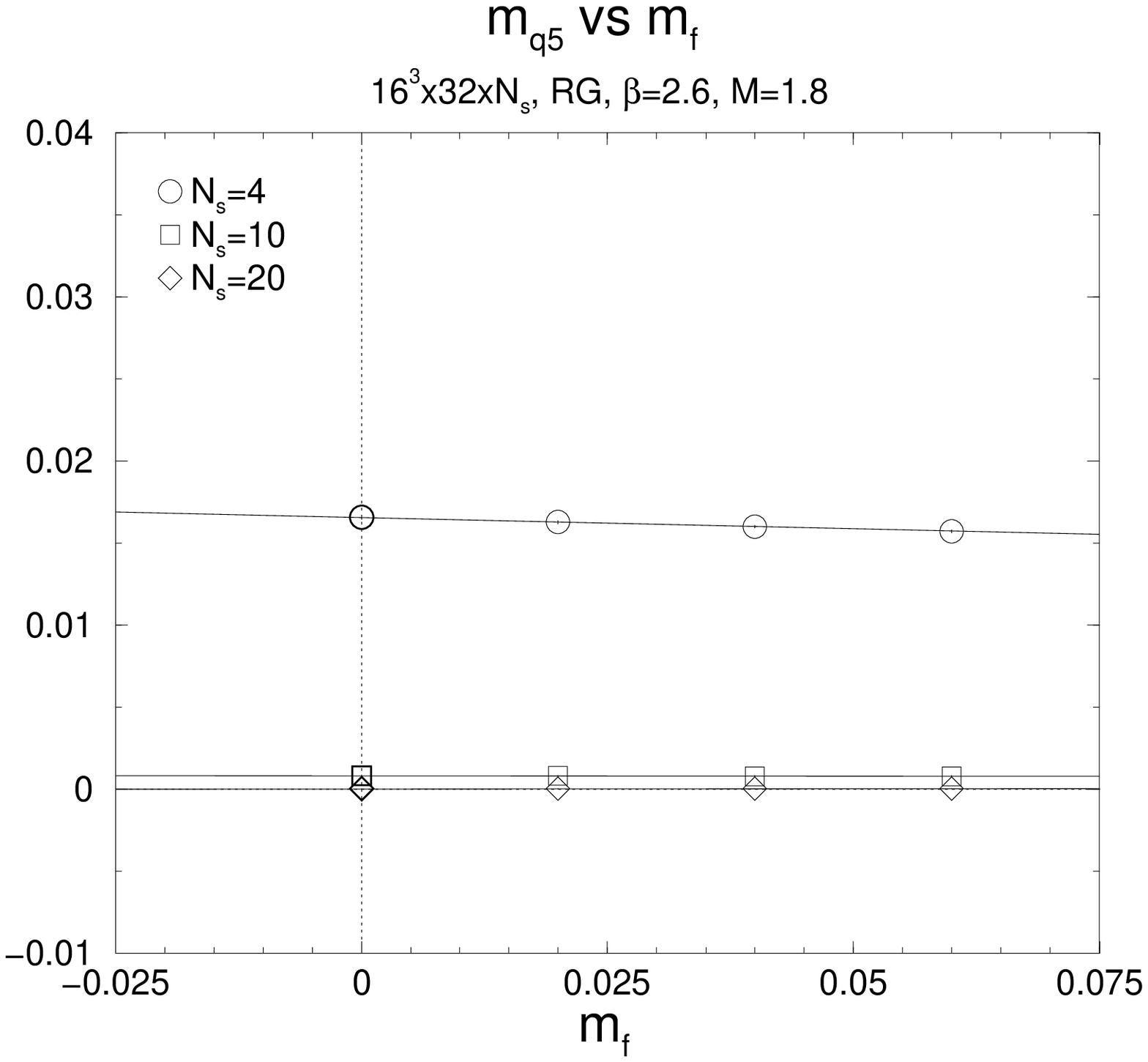}}
\vspace{-1.0cm}
\caption{$m_{q5}$ as a function of $m_f$ 
on a $16^3\times 32 \times N_s$ ($N_s =4$,10 and 20)
at $M=1.8$ and $\beta = 2.6$ 
for the renormalization group (RG) improved gauge action.}
\label{fig:mq5mf}
\vspace{-0.5cm}
\end{figure}

As suggested by the behavior of the pole mass,
the DWQCD satisfies the almost exact axial Ward-Takahashi identity,
which is given by
\begin{eqnarray}
& &\langle \left\{ \triangle_4 A_4^b (t) + 2 m_f \bar q\gamma_5\tau^b q (t)
- 2 {j_5^b (t, N_s/2)}\right\} {\cal O} \rangle \nn \\
& & + \langle \delta^b_t {\cal O} \rangle = 0
\end{eqnarray}
where
$A_4^b $ is an axial-vector current,
$\cal O$ is an arbitrary operators which contain $q$ and $\bar q$ only
as the fermion variable,
and $j_5^b (t, N_s/2)$, defined in the middle of the 5-th dimension,
represents the  explicit breaking term of the chiral symmetry.

In perturbation theory
it is proven that $\langle j_5^b (t, N_s/2) {\cal O} \rangle \rightarrow 0$
as $N_s\rightarrow\infty $ for the non-singlet sector($b\not= 0$),
while $\langle j_5^b (t, N_s/2) {\cal O} \rangle \propto
F_{\mu\nu} \tilde{F}_{\mu\nu} \not= 0$ 
for the singlet sector ($b=0$).
The latter one corresponds to the U$_A$(1) anomaly,
well-known in the continuum QCD.

Again the presence of the exact axial Ward-Takahashi in the DWQCD
suggests the existence of the chiral symmetry in the
$N_s \rightarrow \infty$ limit even for interacting theories.

\section{Numerical investigation of chiral properties of the DWQCD}
\begin{figure}[tb]
\vspace{-0.5cm}
\centerline{\epsfxsize=6.0cm \epsfbox{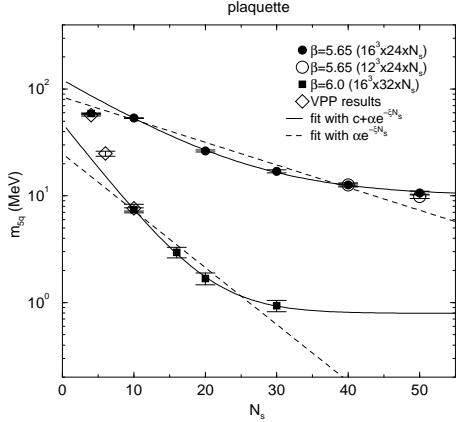}}
\vspace{-1.0cm}
\caption{$m_{q5}$ as a function of $N_s$ for the plaquette gauge action
at $a\simeq 0.2$ fm(circles) and 0.1 fm(squares).}
\label{fig:mq5NsP}
\vspace{-0.5cm}
\end{figure}

In order to investigate the chiral properties of the DWQCD,
we define the anomalous quark mass in axial Ward-Takahashi identity
as follows.
\begin{eqnarray}
m_{q5} &=& \lim_{t\rightarrow\infty}
\frac{\langle j_5^a(t,N_s/2)\cdot \bar q\gamma_5 \tau^aq (0) \rangle}
{\langle  \bar q\gamma_5 \tau^a q (t) \cdot\bar q\gamma_5 \tau^a q (0) \rangle}
\end{eqnarray}
We expect that $m_{q5}$ behaves as
\begin{eqnarray}
m_{q5} &=&  C + \alpha \exp [ -\xi N_s]+ \gamma m_f  + O(m_f^2) .
\end{eqnarray}
The existence of the massless fermion requires $C=0$.

In Fig.~\ref{fig:mq5mf}, we have plotted 
$m_{q5}$ as a function of $m_f$ with $M=1.8$ on a
$L^3\times T \times N_s = 16^3\times 32 \times (4,10,20)$ lattice
at $\beta = 2.6$ of the renormalization group (RG) improved gauge action
\cite{cppacs-dwf}.

The figure shows that 
$m_f$ dependence of $m_{q5}$ is much milder than that of $m_\pi^2$
and is almost constant, so that the extrapolation of $m_{q5}$
to $m_f=0$ becomes much easier.
In addition $m_{q5}$ is very precise and sensitive to $N_s$,
so that the $N_s$ dependence can be extracted.
Hereafter we exclusively employ $m_{q5}$ to investigate
the chiral properties of the DWQCD, because of these advantages of
$m_{q5}$.

\begin{figure}[tb]
\vspace{-0.5cm}
\centerline{\epsfxsize=6.0cm \epsfbox{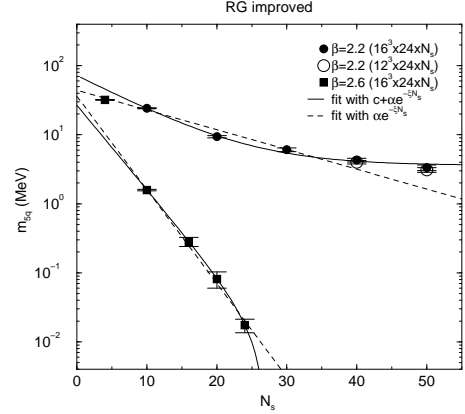}}
\vspace{-1.0cm}
\caption{Same as Fig.\protect{\ref{fig:mq5NsP}} for the RG improved
gauge action.}
\label{fig:mq5NsR}
\vspace{-0.5cm}
\end{figure}

We have plotted $m_{q5}$ as a function of $N_s$,
calculated on the CP-PACS 
for the plaquette gauge action in Fig.\ref{fig:mq5NsP} and 
the RG improved gauge action(lower) in Fig.\ref{fig:mq5NsR},
at coarse($a\simeq 0.2$ fm) and  fine($a\simeq 0.1$ fm) lattice 
spacings\cite{cppacs-dwf}.

At both lattice spacings
$m_{q5}$ is smaller for the RG improved action than for the plaquette action.
At the coarse lattice spacing($a \simeq$ 0.2 fm) it seems that
$C \not= 0$ for both actions. On the other hand,
$C = 0$ for the RG action while $C \not= 0$ for the plaquette action
at the fine lattice spacing ($a \simeq$ 0.1 fm).

\section{Phase structure of the Wilson fermion and the DWQCD}
In this section we try to understand the dependences of $m_{q5}$
on the lattice spacings and the gauge actions.
The existence of massless fermion of the DWQCD can be proven in
the absence of zero eigenvalues of the hermitian
Wilson-Dirac operator $H_W =\gamma_5 D_W(M)$, where $D_W(M)$ is the
4-dimensional Wilson fermion operator with the fermion mass $M$ and
$D_W(M)^\dagger = \gamma_5 D_W(M)\gamma_5$ is satisfied.

In the case of the 4-dimensional Wilson fermion,
zero eigenvalues of $\gamma_5 D_W$  leads to the
spontaneous breaking of parity-flavor symmetries,
$\langle \bar q i\gamma_5 \tau^3 q\rangle \not=0$,
as shown below\cite{Aoki-phase}.

\begin{eqnarray}
& &\langle \bar q i\gamma_5 \tau^3 q\rangle =
-\lim_{H\rightarrow+0} {\rm Tr}\frac{i\gamma_5\tau^3}{D_W+i\gamma_5\tau^3 H} 
\nn \\
&=&-\lim_{H\rightarrow+0} {\rm tr}\left[
\frac{i\gamma_5}{D_W+i\gamma_5 H} - \frac{i\gamma_5}{D_W-i\gamma_5 H} \right]
\nn\\
&=&-i\lim_{H\rightarrow+0} {\rm tr}
\left[\frac{1}{H_W+i H} - \frac{1}{H_W-i H} \right] \nn \\
&=&-i\lim_{H\rightarrow +0}\sum_n 
\VEV{\lambda_n}{\left(\frac{1}{\lambda_n+i H} - \frac{1}{\lambda_n-i H}
\right)}{\lambda_n}\nn \\
&=& -i\lim_{H\rightarrow+0}\int d\,\lambda \, \rho_{H_W}(\lambda)
\VEV{\lambda}{\left(\frac{1}{\lambda+i H} - \frac{1}{\lambda-i H}
\right)}{\lambda} \nn \\
&=&-i \int d\,\lambda \,\rho_{H_W}(\lambda) (-2\pi i)\delta(\lambda)\nn\\
&=& -2\pi\,\rho_{H_W}(0) 
\end{eqnarray}
where 
$\rho_{H_W}(\lambda)$ is the density of the eigenvalues of $H_W$, which is 
defined by
\[
\rho_{H_W}(\lambda)=\sum_n \delta(\lambda_n-\lambda).
\]

\begin{figure}[tb]
\vspace{-0.5cm}
\centerline{\epsfxsize=6.0cm \epsfbox{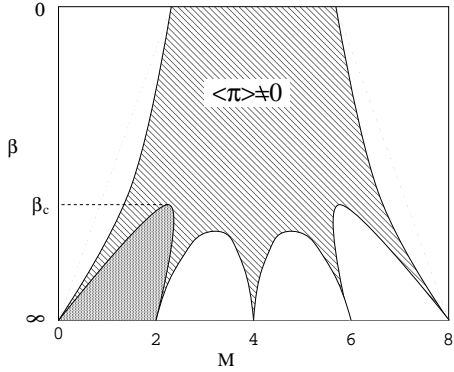}}
\vspace{-1.0cm}
\caption{The expected phase structure of the lattice QCD with the Wilson
fermion in the $\beta=6/g^2$ --$M$ plane,
where $\pi = \bar q i\gamma_5\tau^3 q$. }
\label{fig:aoki_phase}
\vspace{-0.5cm}
\end{figure}

In Fig.\ref{fig:aoki_phase}, 
we have drawn the expected phase structure of the lattice QCD with the
Wilson fermion in the $\beta =6/g^2$--$M$ plane, where
$g^2$ is the gauge coupling constant\cite{Aoki-phase}. 
The small value of $\beta$ corresponds to the coarse lattice spacing
while the large value to the fine lattice.
At $g^2=0$ ($\beta =\infty$) no zero eigenvalues exist for $0 < M < 2$,
which is the region where the domain-wall fermion has the zero modes.
If we increase the gauge coupling, the allowed region of $M$ 
for the massless DWF becomes narrow and is shifted to larger values of $M$.
Finally the allowed region disappears at $\beta =\beta_c$.
This implies that 
massless fermions disappear  at $\beta < \beta_c$
for the (quenched) DWQCD.

According to this interpretation
it seems that
$\beta_c > 5.65$ ($a=0.2$ fm),  6.0($a=0.1$ fm) for 
the plaquette gauge action, while the numerical data indicate
that 2.6 ($a=0.1$ fm) $ >\beta_c >$ 2.2 ($a=0.2$ fm) for the RG improved
gauge action.

\section{Eigenvalues of $H_W$ and $\rho_{H_W}(0)$}
In this section we directly investigate the distribution
of small eigenvalues of $H_W$.

\begin{figure}[t]
\centerline{\epsfxsize=7.5cm \epsfbox{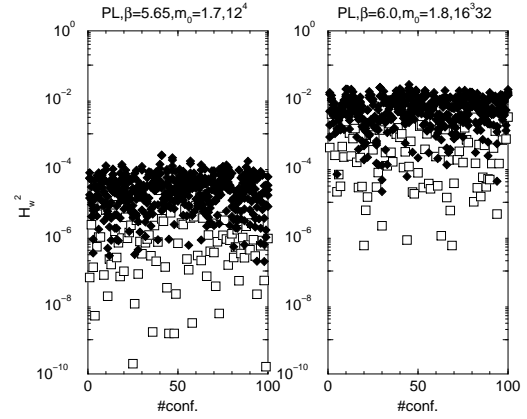}}
\vspace{-1cm}
\caption{
Monte Carlo time histories for the lowest six eigenvalues of
$H_W^2$ obtained with the plaquette gauge action.
}
\label{fig:evdistP}
\vspace{-0.5cm}
\end{figure}

\begin{figure}[t]
\centerline{\epsfxsize=7.5cm \epsfbox{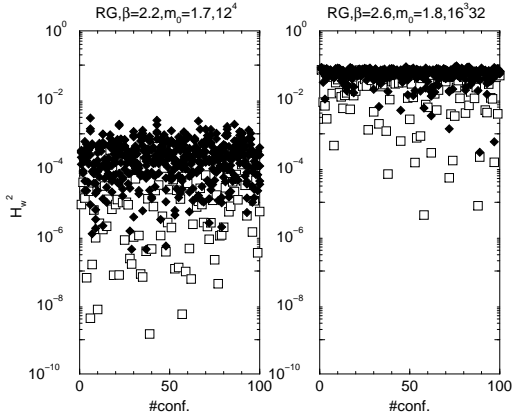}}
\vspace{-1cm}
\caption{
The same as Fig.~\protect\ref{fig:evdistP} obtained with the RG-improved
gauge action.
}
\label{fig:evdistR}
\vspace{-0.5cm}
\end{figure}

In Figs.~\ref{fig:evdistP} and \ref{fig:evdistR}
we plot Monte Carlo time histories for the lowest six eigenvalues of
$H_W^2$ for the plaquette and RG-improved 
actions\cite{cppacs-nagai}.
The figures at $a \simeq 0.1$~fm are shown
at the same lattice size with the previous work of $m_{5q}$.
In each figure the left panel shows results for $a\simeq 0.2$~fm
and the right panel for $a \simeq 0.1$~fm.
Open squares show the minimum eigenvalue $\lambda^2_{\rm min}$ and
filled diamonds are the higher five eigenvalues.

There is a clear trend that the minimum eigenvalues become larger
for smaller lattice spacings.  Another interesting point is that
the RG-improved action gives larger values $\lambda^2_{\rm min}$ than the
plaquette action, which indicates that the RG-improved action has a
better chiral behavior.

The spectral density of $H_W$ is defined by
\begin{equation}
\rho_{H_W}(\lambda) = \lim_{V \rightarrow \infty}
\frac{1}{3\cdot 4\cdot V}\sum_{\lambda'}\delta (\lambda'-\lambda),
\end{equation}
where the summation is over the eigenvalues of $H_W$.
The density of zero-eigenvalues $\rho_{H_W}(0)$, related to the
order parameter of the parity-flavor breaking,
has been calculated by
the accumulation method proposed in \cite{EHN99},
which is based on the relation
\begin{eqnarray}
\frac{1}{3 \cdot 4 \cdot V}
\sum_{|\lambda^\prime| \le \lambda}
\bf{1}& = & \int_{-\lambda}^\lambda d \lambda^\prime \rho(\lambda^\prime)
\nn \\
&\simeq& 2 \rho(0) \lambda  + O(\lambda^2).
\label{eq:ansatz}
\end{eqnarray}
We note that, for the small-$\lambda$ expansion 
in (\ref{eq:ansatz}),
analyticity of $\rho(\lambda)$ at the origin is assumed.

\begin{figure}[tb]
\vspace{-0.5cm}
\centerline{\epsfxsize=6.0cm \epsfbox{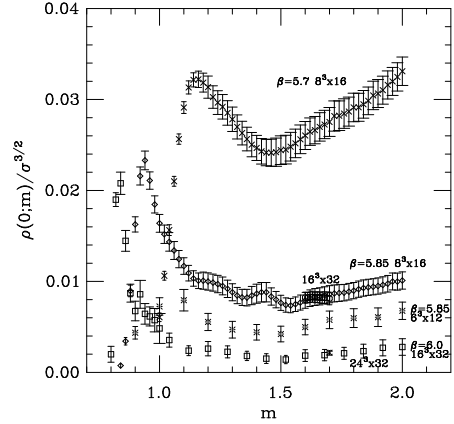}}
\vspace{-1.0cm}
\caption{$\rho_{H_W}(0)$ as a function of $M$ at several values of $\beta$
for the plaquette action\protect{\cite{EHN99}}.}
\label{fig:rhoM}
\vspace{-0.5cm}
\end{figure}

\begin{figure}[tb]
\vspace{-0.5cm}
\centerline{\epsfxsize=6.0cm \epsfbox{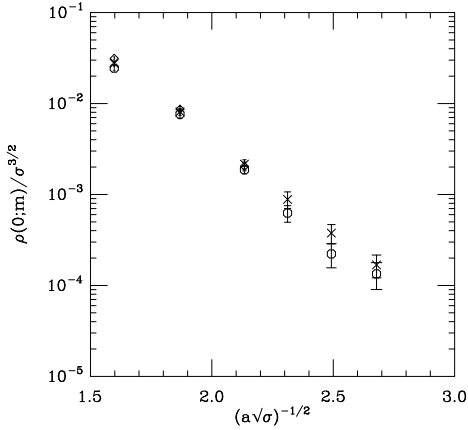}}
\vspace{-1.0cm}
\caption{$\rho_{H_W}(0)$ as a function of $a^{-1/2}$ at $M=1.2$ for
the plaquette action\protect{\cite{EHN99}}.}
\label{fig:rho-a}
\vspace{-0.5cm}
\end{figure}

The result of \cite{EHN99}, where $\rho_{H_W}(0)$ at several $\beta$'s
have been plotted as a function of $M$, is given in Fig.~\ref{fig:rhoM}
for the plaquette gauge action.
Although $\rho_{H_W}(0)$ becomes smaller for smaller $a$
(; larger $\beta$), no region without zero eigenvalues
seems to exist.
This impression is indeed confirmed in Fig.~\ref{fig:rho-a},
where $\rho_{H_W}(0)$ has been plotted as a function of
$ a^{-1/2}$ at $M=1.2$\cite{EHN99}.

It has been speculated in \cite{EHN} that the $a$ dependence of
$\rho_{H_W}(0)$ is well reproduced by the from that
\begin{equation}
\rho_{H_W}(0) \simeq A \exp[- \frac{c}{ \sqrt{ a }}] .
\end{equation}
If this form is correct for all $M$, the region with
$\rho_{H_W}(0)= 0$ does not exist as long as $a\not= 0$.
The similar result has been obtained for the RG improved action.

These results immediately lead to the conclusion  that there is
no gap of the parity-flavor breaking phase
in the $g^2$--$M$ plane for the Wilson fermion, so that
the domain-wall QCD fails to realize the chiral symmetry
at all $a\not= 0$.
However this conclusion seems to contradict the numerical data
for the RG improved gauge action at $a = 0.1$ fm.
Puzzles still remain.

\section{Theoretical understanding for the behavior of $m_{q5}$}
In this section we try to understand the pathological behavior of
$m_{q5}$ as a function of $N_s$, in terms of the eigenvalues of $H_W$.

We first derive the analytic expression for $m_{q5}$ as follows.
\begin{equation}
\frac{2 m_{q5} a_5}{(1-m_f)^2}=
\frac{\sum_{X,Y}\vert \VEV{X}{f(\tilde H)^{-1} H_{GW}^{-1}}{Y}\vert^2}
{\sum_{X,Y}\vert \VEV{X}{ H_{GW}^{-1}}{Y}\vert^2}
\end{equation}
where $X =(x,\alpha,a)$ represents site, spinor
and color indices,
\begin{eqnarray*}
f(\tilde H) &=&\cosh( \tilde H N_s/2) =\frac{T^{-N_s/2}+T^{N_s/2}}{2}\\
T &=& \frac{1-H_W^\prime}{1+H_W^\prime}\\
H_W^\prime &=&H_W \frac{1}{2+\gamma_5 H_W} \\
H_{GW}^{-1} &=& \frac{1}{(1+m_f)+(1-m_f)\gamma_5 \tanh (\tilde H N_s/2)}
\gamma_5
\end{eqnarray*}
For the derivation of the above formula, see \cite{AT}.

Based on the above formula $m_{q5}$ can be approximately estimated
at $m_f=0$ as follows.
\begin{eqnarray}
& &2 m_{q5} a_5 = \frac{1}{N_{all}}
\left[
c_0\sum_{n=1}^{N_D}f(\tilde\lambda_n)^{-2}\right. \nn \\
& &+ \left. (N_{all}-N_D)\int d\lambda\rho(\lambda)
f(\tilde\lambda)^{-2} \right] \nn \\
& & =\frac{1}{N_{all}}\left[
c_0\sum_n f(\tilde\lambda_n)^{-2} + \sum_c f(\tilde\lambda_c)^{-2}
\right] \\
\tilde\lambda &=& \log\frac{1-\lambda}{1+\lambda}
\end{eqnarray}
where $\lambda_c$($\lambda_n$) is the continuous(discrete) 
eigenvalue of $H_W^\prime$,
$N_D$ is the number of discrete modes,
$N_{all}= L^3 \times T\times N_c\times 4$ is the number of degrees of 
freedom, and $c_0$ is a weight factor for the discrete modes.
See \cite{AT} for the detail of the estimation.

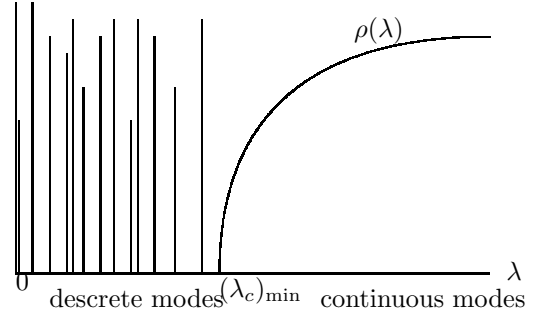
\begin{figure}[tb]
\vspace{-0.5cm}
\begin{center}
\setlength{\unitlength}{0.45mm}
\begin{picture}(150,90)
\put(0,10){\line(0,1){80}}
\put(0,10){\line(1,0){140}}
{\bezier{800}(60,10)(60,80)(140,80)}
\put(100,80){$\rho(\lambda)$}
\put(1,10){\line(0,1){45}}
\put(5,10){\line(0,1){80}}
\put(10,10){\line(0,1){70}}
\put(15,10){\line(0,1){65}}
\put(17,10){\line(0,1){75}}
\put(20,10){\line(0,1){55}}
\put(25,10){\line(0,1){70}}
\put(29,10){\line(0,1){75}}
\put(34,10){\line(0,1){45}}
\put(36,10){\line(0,1){75}}
\put(41,10){\line(0,1){70}}
\put(47,10){\line(0,1){55}}
\put(55,10){\line(0,1){75}}
\put(0,5){0}
\put(145,8){$\lambda$}
\put(10,0){descrete modes}
\put(90,0){continuous modes}
\put(60,3){$(\lambda_c)_{\rm min}$}
\end{picture}
\end{center}
\vspace{-1.0cm}
\caption{The expected distribution of the eigenvalues of $\tilde H_W$
in the weak coupling region.}
\label{fig:dist}
\vspace{-0.5cm}
\end{figure}

The important point is that there are two types of eigenvalues,
the continuous eigenvalues and the discrete eigenvalues.
The eigenfunction for the former is the plain-wave like and spreads
over whole space, while the one for the latter is exponentially localized.
In the strong coupling region at $\beta < \beta_c$,
the continuous eigenvalues can become zero, so that
$\rho_{H_W}(0)\not=0$. Therefore the DWQCD does not work as expected.
(See \cite{AK} for more details.)
In the weak coupling region, on the other hand,
the continuous eigenvalues has a minimum value $(\lambda_c)_{\rm min}$,
such that $\rho_{H_W}(\lambda_c < (\lambda_c)_{\rm min}) = 0$
for the continuous eigenvalues.
Instead of the continuous eigenvalues,
discrete eigenvalues may become almost zero.
The expected distribution of eigenvalues $\lambda$ in the
weak coupling region is schematically drawn
in Fig.~\ref{fig:dist}.

It is likely that small (localized) discrete eigenvalues
are caused by dislocations. In particular,
$Q$, the value of the topological charge in some configuration
should move to the different value $Q^\prime$ during the simulation of
the lattice QCD, so that the dislocation appears during the transition
from $Q$ to $Q^\prime$.
If the dislocation appears, the Wilson-Dirac operator has an almost
zero eigenvalue at some value of $M$.
Therefore it is very difficult to avoid small (localized) eigenvalues
of $H_W$ and $\tilde H_W$.

From the approximated estimate for $m_{q5}$, only near zero
eigenvalues can contribute to it as $N_s\rightarrow\infty$.
Hence only small discrete eigenvalues are dangerous for the
DWQCD in the weak coupling region.
The analysis in \cite{AT} leads to the followings.

In the $N_s\rightarrow\infty$ limit 
$m_{q5}$ vanish at finite volume (;$N_{all}\not= \infty$), 
since the probability of having exact zero eigenvalues is zero.

If the infinite volume (;$N_{all}\rightarrow \infty$) limit 
is taken before the $N_s\rightarrow\infty$ limit,
there are two possibilities for the distribution of discrete
eigenvalues.
Unless the number of near zero eigenvalues increases linearly in
$N_{all}$, they does not contribute to $m_{q5}$, hence the DWQCD
works fine.
If it increases linearly in $N_{all}$, $\rho_{H_W}(0)$ and hence
$m_{q5}$ becomes non-zero, so that 
the DWQCD can not realize the chiral symmetry.

The numerical data obtained so far seem to prefer the latter possibility.
However more detailed investigations will be necessary for the definite
conclusion.

\section{Conclusion}
The distribution of near zero eigenvalues of $H_W$ plays a crucial role
for the chiral properties of the DWQCD.
We have pointed out that there are two types of eigenvalues,
continuous and discrete.
In the weak coupling region, only discrete eigenvalues become relevant
for $m_{q5}$. However the effect of discrete eigenvalues to
$m_{q5}$ is rather small even if the number of near zero
discrete eigenvalues increases linearly in the volume.
We think that this has brought confusing numerical results for the 
$N_s$ dependence of $m_{q5}$.

Our analysis suggests that the gap is closed in the phase structure of the
Wilson fermion due to the near zero discrete eigenvalues,
if the quenched approximation is employed.
In the dynamical QCD simulation, the near zero discrete eigenvalues are
strongly suppressed by the fermion determinant,
so that the gap opens and the expected phase structure is correct
in the real lattice QCD.
Since such a suppression of small discrete eigenvalues is not known for
the DWQCD, it is unclear how the dynamical DWQCD change the 
distribution of the near zero discrete eigenvalues.

\section*{Acknowledgments}
I would like to thank Dr. Y.~Taniguchi, with whom I
have completed the main part of this work.
I also thank the members of the CP-PACS collaboration for
some results presented in this report, and
Prof.~H.~Neuberger for the useful discussion.
This work is supported in part by Grants-in-Aid
of the Ministry of Education(No. 12640253).

\newcommand{\J}[4]{{#1} {#2}, #3 (#4)}
\newcommand{\AP}{Ann.~Phys.}
\newcommand{\CMP}{Commun.~Math.~Phys.}
\newcommand{\IJMP}{Int.~J.~Mod.~Phys.}
\newcommand{\MPL}{Mod.~Phys.~Lett.}
\newcommand{\NP}{Nucl.~Phys.}
\newcommand{\NPSup}{Nucl.~Phys.~B (Proc.~Suppl.)}
\newcommand{\PL}{Phys.~Lett.}
\newcommand{\PR}{Phys.~Rev.}
\newcommand{\PRL}{Phys.~Rev.~Lett.}
\newcommand{\PTP}{Prog. Theor. Phys.}
\newcommand{\Suppl}{Prog. Theor. Phys. Suppl.}
\newcommand{\RMP}{Rev. Mod. Phys.}


\begin{thebibliography}{99}

\bibitem{GW}
P. Ginsparg and K. Wilson,
\J{\PR}{D25}{2649}{1982}.

\bibitem{Luscher98}
M.~L\"uscher,
 \J{\PL}{B428}{342}{1998}.

\bibitem{perfect1}
P.~Hasenfratz,
 \J{\NP}{B525}{401}{1998};
P.~Hasenfratz, V.~Laliena and F.~Niedermayer,
 \J{\PL}{B427}{125}{1998}.

\bibitem{Neuberger98}
H.~Neuberger,
 \J{\PL}{B417}{141}{1998}; \J{\PL}{B427}{353}{1998};
 \J{\PR}{D57}{5417}{1998}.

\bibitem{NN94}
R.~Narayanan and H.~Neuberger,
\J{\NP}{B412}{574}{1994}; \J{\NP}{B443}{305}{1995}.

\bibitem{Kaplan92}
D.~Kaplan,  \J{\PL}{B288}{342}{1992}.

\bibitem{Shamir93}
Y.~Shamir,  \J{\NP}{B406}{90}{1993};
V.~Furman and Y.~Shamir,
 \J{\NP}{B439}{54}{1995}.

\bibitem{Blum-Soni} T.~Blum and A.~Soni,
 \J{\PR}{D56}{174}{1997}; \J{\PRL}{79}{3595}{1997}; hep-lat/9712004.

\bibitem{AIKT}
S.~Aoki, T.~Izubuchi, Y.~Kuramashi and Y.~Taniguchi,
\J{\PR}{D62}{094502}{2000}.

\bibitem{Blum98}
For a review, see
T.~Blum,
 \J{\NPSup}{73}{167}{1999}
and references there in.

\bibitem{cppacs-dwf}
CP-PACS Collaboration, A. Ali Khan {\it et al.},
\J{\PR}{D63}{114504}{2001}; \J{\NPSup}{83-84}{591}{2000}.

\bibitem{HJL98}
P.~Hernandez, K.~Jansen and M.~L\"uscher,
\J{\NP}{B552}{363}{1999}.

\bibitem{KN99}
Y.~Kikukawa,
\J{\NP}{B584}{511}{2000};
Y.~Kikukawa and T.~Noguchi, hep-lat/9902022.

\bibitem{Aoki-phase}
S.~Aoki,
 \J{\PR}{D30}{2653}{1984}; \J{\PRL}{57}{3136}{1986}; \J{\NP}{B314}{79}{1989}.

\bibitem{EHN99}
R. Edwards, U. Heller and R. Narayanan,
\J{\PR}{D60}{1999}{034502}.

\bibitem{cppacs-nagai}
CP-PACS Collaboration, A. Ali Khan {\it et al.},
\J{\NPSup}{94}{725}{2001}

\bibitem{AT}
S.~Aoki and Y.~Taniguchi,
hep-lat/0109022.

\end{thebibliography}
\end{document}